\documentclass[conference]{IEEEtran}
\IEEEoverridecommandlockouts
\usepackage{cite}
\usepackage{amsmath,amssymb,amsfonts}
\usepackage{algorithmic}
\usepackage{graphicx}
\usepackage{textcomp}
\usepackage{xcolor}
\usepackage{graphicx}
\usepackage{float}
\usepackage{booktabs}
\usepackage{geometry}

\newcommand{\CLASSINPUTtoptextmargin}{0.75in}
\def\BibTeX{{\rm B\kern-.05em{\sc i\kern-.025em b}\kern-.08em
    T\kern-.1667em\lower.7ex\hbox{E}\kern-.125emX}}
\begin{document}

\title{Securing UAV Communication : Authentication and Integrity
}

\author{
  \IEEEauthorblockN{Meriem Ouadah }
  \IEEEauthorblockA{
    \textit{Telecommunications Department}\\
   \textit{LISIC Laboratory, USTHB University}\\
    Algiers, Algeria\\
    mouadah@usthb.dz
  }
  \and
  \and
  \IEEEauthorblockN{  Fatiha Merazka }
  \IEEEauthorblockA{
    \textit{Telecommunications Department}\\
   \textit{LISIC Laboratory, USTHB University}\\
    Algiers, Algeria\\
    fmerazka@usthb.dz 
  }
}
\IEEEoverridecommandlockouts
\IEEEoverridecommandlockouts\IEEEpubid{\makebox[\columnwidth]{979-8-3503-7786-6/24/\$31.00~\copyright~2024~IEEE } \hspace{\columnsep}\makebox[\columnwidth]{ }}
\maketitle

\begin{abstract}
Recent technological advancements have seen the integration of unmanned aerial networks (UAVs) into various sectors, from civilian missions to military operations. In this context, ensuring security, precisely authentication, is essential to prevent data theft and manipulation. A Man-in-the-Middle attack not only compromises network integrity but also threatens the original data, potentially leading to theft or alteration.
In this work, we proposed an authentication method to secure UAV data exchange over an insecure communication channel. Our solution combines Diffie-Hellman (DH) key exchange and Hash-based Message Authentication Code (HMAC) within ROS communication channels to authenticate exchanged UAV data. We evaluated our method by measuring transmission time and simulating key tampering, finding acceptable performance for DH key sizes below 4096 bits but longer times for larger sizes due to increased complexity. Both drones successfully detected tampered keys, affirming our method's efficacy in protecting UAV communication. However, scalability challenges in resource-constrained environments warrant further research.
\end{abstract}


\section{Introduction}
Widely recognized as drones, unmanned aerial vehicles (UAVs) networks have attracted a lot of research attention in the past few years as they facilitate network communication. As indicated by the name, UAVs are aircrafts that are automatically piloted using computers instead of having a human pilot.
UAVs belong to the non-terrestrial network category as they move from land to space providing many advantages unlike terrestrial networks. First, thanks to their mobility, they provide enhanced connectivity by easily moving in the environment allowing network connection to reach the less fortunate rural areas where people do not have optical fibers and can only use 3G networks \cite{b1}. In addition to that, various algorithms were also proposed to optimize and increase mobility using factors such as coverage maximization and power control \cite{b2} and the optimal 3D trajectory of each UAV \cite{b3}.
Secondly, UAVs enhance energy efficiency, particularly for Internet of Things devices that suffer from limited battery capacity. Acting as relays between sensors in remote and urban areas that have energy shortages, UAVs collect the IoT data to ease their transmission to unreachable areas \cite{b4}.

The previously discussed features position UAVs as a perfect candidate for integration with beyond 5G and 6G networks where millions of interconnected devices are communicating with each other requiring very high throughput, low latency, connectivity, and energy \cite{b5}.  However,ensuring the security of these networks is crucial to create a complete system. Indeed, security remains a significant challenge in futuristic non-terrestrial communication networks, often overlooked in the development of UAV intelligent applications given their constrained computing capabilities \cite{b6}. Prioritizing 6G security is crucial, with concepts such as blockchain, quantum computing and communication technology, security protocols, and others being considered to enhance resilience against cyberattacks \cite{b7}.

UAV applications
range from using communication between UAVs and base
stations to novel approaches utilizing a swarm network of
UAVs and employing UAV-to-UAV communication to create
an interactive environment where data and/or communication
is exchanged between these devices.

According to this, various drone networking security solutions have been proposed by researchers. In this article, we focus on two of the security triad components which are authentication and integrity by proposing a secure communication approach that combines the Diffie-Hellman key exchange method for authentication and the use of the HMAC algorithm for integrity.
Our contribution is highlighted in three aspects: 
\begin{itemize}
\item The simplicity and flexibility of the authentication method make it well-suited for UAV communication environments.
\item The proposed method adds another security layer that suits the low energy capacity of UAVs.
\item Robustness is enhanced by leveraging ROS (Robotic Operating System) communication channels, which provide standardized and reliable communication for UAV systems. 
\end{itemize}
The rest of the paper is organized as follows: section II presents related works and research solutions in the field of drone security as well as innovations in UAV architecture designs. Section III describes the proposed system, its architecture, and the simulation environment. In section IV, we discuss the simulation results and the performance evaluation. Finally, we conclude and present future works.

\section{Related work}

A work in \cite{b8} addressed the security of an Internet of Drone environment (IoD) where a drone is traveling a long distance to deliver goods to the designated customer location and then must return safely escaping all network attacks. The proposed system uses the asymmetric Elliptic-curse-cryptography (ECC) algorithm to encrypt the sent data, it uses a shared key for encryption and a secret one for decryption, furthermore, the combination of symmetric and asymmetric encryption is proposed to increase security, a hybrid encryption and decryption mechanism are realized at the transmitter and receiver nodes respectively \cite{b8}. Results show that the data sent from attackers acting as fake nodes are verified by SHA-256 and easily ignored \cite{b8}. Researchers in \cite{b9} divided UAV security into 03 main categories:
\begin{itemize}
\item Protocol-based security: Including UranusLink protocol that identifies packet loss using sequence numbers to optimize UAV long-distance communication and improve link performance by up to 33\%, UAVCAN that improves communication abstraction to support diverse applications by enhancing the medium support, and MavLink protocol, which is used for bidirectional communication between UAVs and base stations with a secondary security layer to protect against attacks like DDoS and eavesdropping.
\item Agent-based security: two methods are discussed, the self-protective UAV Network method (ASP-UAS) using the Hierarchical Immune System (HIS) and multi-agent systems that enhance security against drone attacks and the HFA algorithm that ensures cybersecurity defense, focusing on secure communication through trust-building and node removal.
\item Trajectory-design-based security: employing deep reinforcement learning, UAV trajectory, and sub-carrier power allocation’s performance has improved by 10.29\% thanks to using a deep Q-network algorithm.
\end{itemize}
Furthermore, in UAV-to-vehicle communication, authors in \cite{b9} studied a case where a UAV is acting as a base station trying to communicate with a vehicle on the highway with the existence of another malicious vehicle on the same highway to eavesdrop on the communication between them. Results are shown below:
\begin{itemize}
\item High influence: has a weak influence on downlink secrecy outage and a minimal impact on uplink performance.
\item Transmit SNR: has a weak impact on downlink secrecy outage and does not significantly affect uplink performance.
\item Ground Coverage Radius: negatively affects secrecy outage in both uplink and downlink scenarios while the radius of the destination's ground coverage has minimal influence.
\end{itemize}
Network security fundamentals are summarized in the CIA acronym, representing Confidentiality, Integrity, and availability. Being the first pillar of the security triad, Confidentiality aims to protect the network from unauthorized access and data theft, it is implemented through access control applications, secure communication protocols, and encryption. According to this, some research works focused on building systems and applications for securing UAV communication by ensuring confidentiality. In [10] a customized Kerberos system was implemented for UAV authentication, it contained a Ground Station, an Authentication Server (AS), and a Ticket Granting Server as a secondary security layer. For the user to communicate with UAVs, they must pass through several phases:
\begin{enumerate}
\item Phase 1: It has two steps, firstly, the user initiates the authentication process by sending a request (message-1) to the Authentication Server (AS) including ID, password, and Ticket-Granting Server (TGS) ID in the secure channel [step 1]. Secondly, the AS verifies the user’s credentials using its database, if authorized, the AS responds (message-2) with the user's private key and an encrypted ticket for TGS access [step 2].
\item Phase 2:  It also has two steps, firstly the user having the TGT, sends a request (message-3) to TGS containing user ID, UAV ID, and TGS ticket [step 3]. Secondly, TGS responds (message-4) with a ticket allowing the user access to the UAV [step 4].
\item Phase 3: Again, two last steps, firstly, the user sends an access request (message-5) to the Ground Station Radio Transceiver, including user ID and TGS ticket, the latter forwards this information (message-6) to the UAV in the Wireless Sensor Network area [step 5]. Secondly, if the user information is correct, the UAV responds with an authorized access message (message-7) to the radio transceiver, the information is then relayed to the user via radio transceiver in message-8 format, containing a notation for UAV access [step 6].
\end{enumerate}
This system resulted in increased speed which is advantageous for enhancing security and defense against intrusion in UAV networks. Additionally, it aims to address shortcomings in power consumption and intrusion for the future \cite{b10}. Another research in \cite{b11} proposed a global and secured authentication system that aims to address UAV communication security challenges by using secure sensor values to defend against attacks and ensure integrity. It is resistant to tampering, especially against remote attacks ensuring availability, authenticity, and confidentiality while being adapted to the acceptance criteria of authorities to attract manufacturers. The system operates against physical and remote attacks by the integration of a Hardware Security Module (HSM) coupled with secure storage for key material, however, it suffers from latency and high power consumption \cite{b11}. Moreover, authors in \cite{b12}proposed a decentralized authentication method to avoid the single point of failure, the system relies on the reliability evaluation of the physical layer fingerprints that divide UAV nodes into selected and unselected nodes, and then a soft authentication judgment is set on the selected ones. When a node whether it was malicious or legitimate tries to send a message to the cluster head (CH), the other node judges if she should be granted access or not \cite{b12}. This method has proven to successfully increase the authentication accuracy while mitigating the wrong estimations that categorize a legitimate UAV as a malicious one and the opposite \cite{b12}.
As a complement to network security, the architectural design must be adapted and well-studied to create a flexible environment for UAV implementation, the SDN technology responds to that by separating the control function from the data layer centralizing the network management, and making the network flexible, programmable and scalable for future implementations \cite{b13}. Several researchers have worked on the integration of UAV-based SDN systems such as SDN-based flying ad hoc network models(FANET), and SDN-based NFV for SDUAV security, others have used this infrastructural combination to prevent collision, reduce end-to-end delay, establish authentication and even in response to COVID-19 challenges \cite{b14}. Over the years more research focused on security for SDN-based UAV networks, for instance, 
We proposed in an earlier research work \cite{b15} a new design that relies on using UAV nodes both as controllers and switches at the same time in SDN, this work consisted of simulating a network pollution attack and using the network coding paradigm (NC) to protect data from being tampered with in case if the attacker gained control over one of the UAV nodes, a verification method of the legitimate IP address was also implemented. While results demonstrated more flexibility and less loss rate, it is still necessary to explore more security methods because IP addresses alone are not enough \cite{b15}.

the difference between our authentication method and those mentioned lies in its simplicity, efficiency, and autonomy. While the Kerberos system introduced in \cite{b10} involves multiple steps and relies on centralized servers, our method operates directly within the UAV network without the need for external servers. Similarly, the authentication system proposed in \cite{b11} may suffer from latency and high power consumption due to its reliance on secure sensor values and hardware security modules. In contrast, our method offers a lightweight solution that balances security requirements with the resource constraints of UAVs, making it more suitable for practical implementation in UAV networks.
While the aforementioned studies contribute significantly to UAV communication security, they primarily focus on specific aspects such as encryption techniques, protocol-based security, and architectural design. In contrast, our proposed method offers a unique approach that addresses multiple security concerns simultaneously including secure authentication and data integrity, safeguarding against unauthorized access, and tampering in UAV communication environments.

\section{System Description}
\begin{figure}[htbp]
\centerline{\includegraphics[width=\columnwidth]{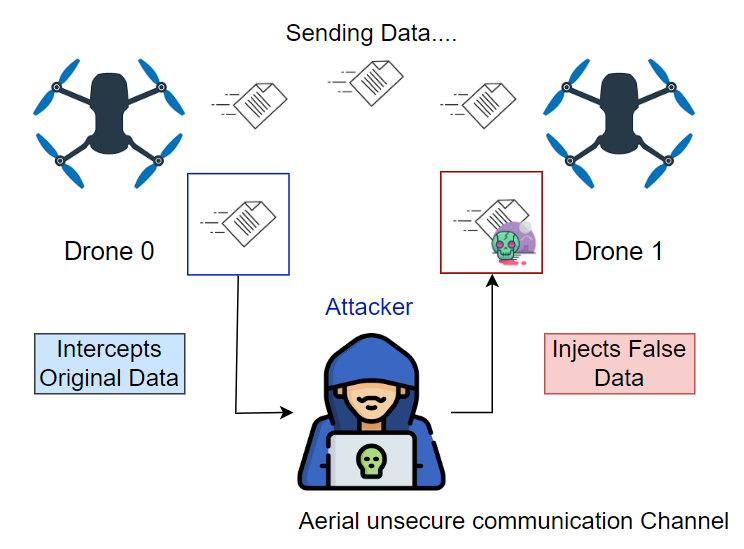}}
\caption{Man in the middle attack in an Inter-UAV communication scenario.}
\label{fig1}
\end{figure}

In an open environment such as the aerial territory in which UAV networks operate, there is no apparent security channel for communicating which makes UAV endangered and exposed to attackers trying to intercept the exchanged information to steal it or manipulate it by changing the original packets. Consequently, research have focused on inter-UAV communication  due to the lack of existing security approaches that protect the inter-communication flow from malicious parties. Fig. \ref{fig1} describes the communication flow from one UAV to another over an insecure communication channel and how Man-In-The-Middle attacks could interfere.

\begin{figure}[htbp]
\centerline{\includegraphics[width=\columnwidth]{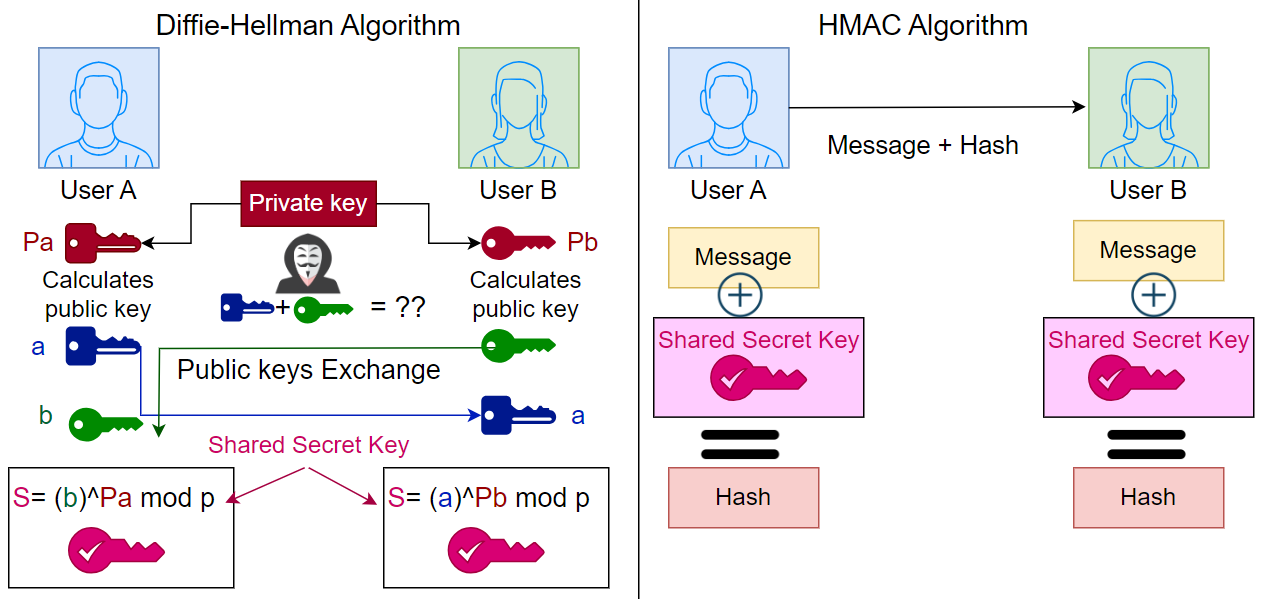}}
\caption{Diffie-Hellman \& HMAC algorithms function.}
\label{fig2}
\end{figure}

According to this, and as one of the actors of the scientific research community, we proposed a solution based on using the combination of the Diffie-Hellman and the HMAC algorithms focusing on applying authentication and integrity in the UAV communication environment. Both algorithms were used in our system to complete the shortenings of each other. Fig. \ref{fig2} illustrates the way they perform.

\begin{enumerate}
\item On the left, we have the Diffie-Hellman algorithm which allows two users (user A and user B) to choose a private key that will not be shared over the network (Pa for A and Pb for B), this key is used in a function to calculate a public key (a for A and b for B), the latter is then exchanged between the two parties and combined with their original private keys to calculate a shared secret key. The used function is described in \ref{fig2} where p is prime. This method prevents an attacker from knowing the shared secret key, even if they intercept the exchanged public keys. 
\item On the right, we have the HMAC algorithm that authenticates the exchanged data and verifies their integrity. When two entities, such as user A and user B want to exchange a message, the message along with a secret key known to both entities only (in our case, the one calculated from Diffie-Hellman) are combined using a hash algorithm to create a hash that will be added to the original message. When user B receives the authenticated data, it will check against the same calculated hash to see whether data has been altered.
\end{enumerate}
\subsection{System Architecture}
\begin{figure}[htbp]
\centerline{\includegraphics[width=\columnwidth]{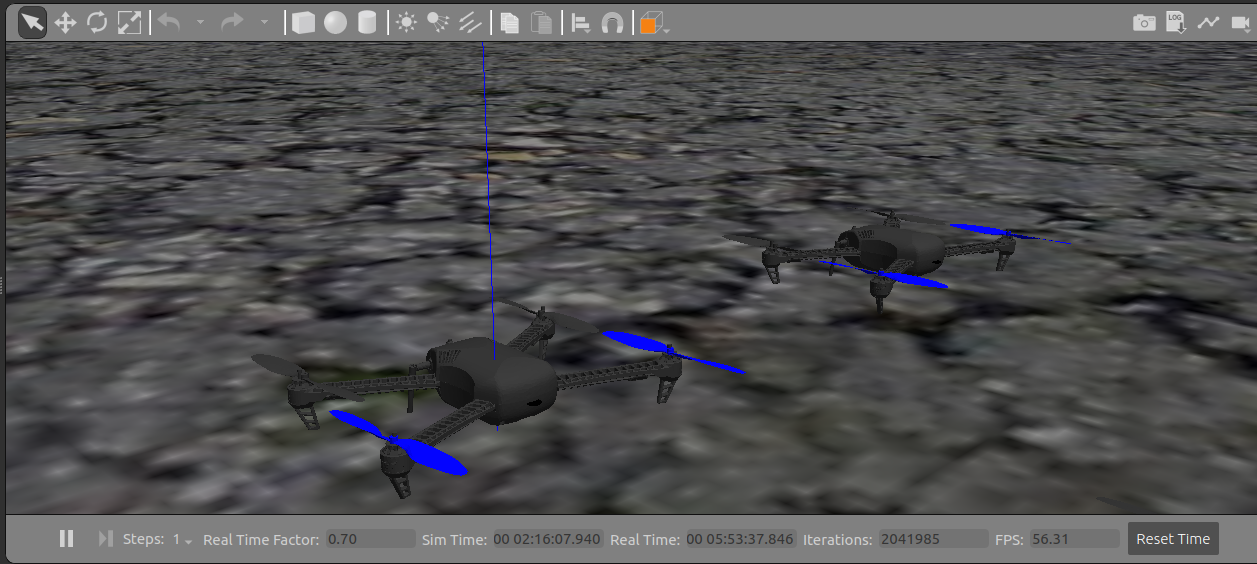}}
\caption{3D Simulation of UAVs over Gazebo interfacing with ROS.}
\label{fig3}
\end{figure}

The architecture focuses on using two drones communicating in a ROS/Gazebo environment as shown in Fig. \ref{fig3}.

The implementation was based on an existing px4-mavros-gazebo-simulation which focuses on establishing a testing and experimentation environment for robotic vehicles like UAVs and Rovers. The four components of this system are explained as follows:
\begin{enumerate}
\item PX4 Autopilot: PX4 is an open-source flight control software designed for unmanned vehicles such as UAVs, drones, and other unmanned vehicles. It offers an all-encompassing platform for the management and navigation control of aerial vehicles. 
\item Gazebo Simulator: Gazebo is a multi-robot simulator that enables the testing and experimenting of robotic algorithms in a simulated setting. It replicates diverse robotic systems by incorporating a realistic physics engine and 3D graphics.
\item MAVROS: MAVROS (Micro Air Vehicle Communication and Control ROS package) is a communication link between the Robot Operating System (ROS) and MAVLink-based flight controllers. It uses the MAVLink communication protocol to streamline the integration of ROS with UAVs.
\item ROS (Robot Operating System): ROS is an open-source middleware framework that aims to control and develop robotic systems. It offers tools and libraries for tasks such as hardware abstraction, device drivers, communication between processes, package management, and others.
\end{enumerate}
\begin{figure}[htbp]
\centerline{\includegraphics[width=\columnwidth]{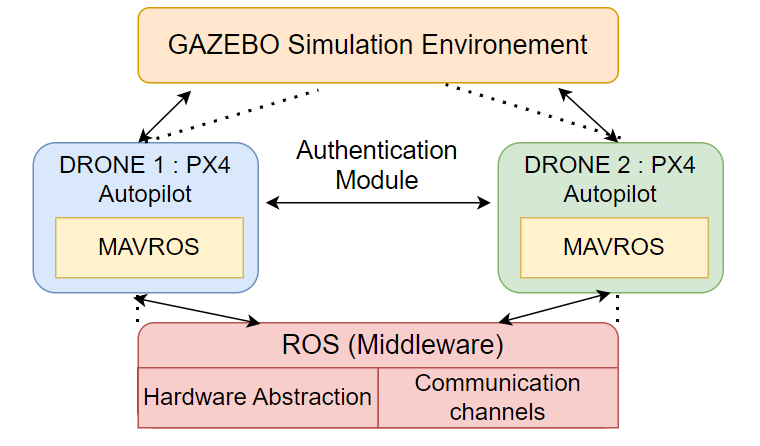}}
\caption{System Architecture.}
\label{fig4}
\end{figure}
Fig. \ref{fig4} illustrates how the four components are related and describes the role of each one of them in our solution. For instance: \begin{itemize} 

\item UAVs operate in a gazebo world.
\item Each drone uses PX4 autopilot for communication and flight control.
\item MAVROS connects PX4 autopilot to ROS and manages communication between UAVs and ROS.
\item ROS enables UAV communication channels, hardware abstraction, and other functionalities.
\item The fifth component is the authentication algorithm that we proposed.
\end{itemize}

\subsection{Proposed Algorithm}
\begin{figure}[htbp]
\centerline{\includegraphics [width=\columnwidth]{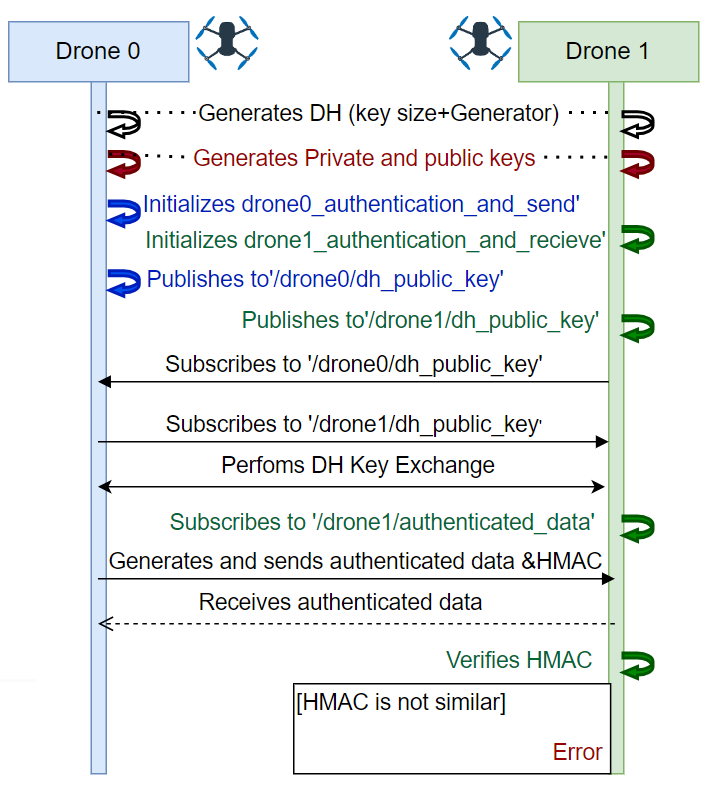}}
\caption{Sequence diagram explaining the chosen algorithm.}
\label{fig5}
\end{figure}

The proposed solution lies in the combination of both the Diffie-Hellman key exchange algorithm and the HMAC algorithm. The latter is used to ensure effective authentication and data integrity due to the well-known Diffie-Hellman weakness against Man in the Middle attacks.
When the key exchange is done and the shared key is calculated, HMAC joins the game by using the secret shared key to calculate its hash and authenticate the exchanged data.
This operation is explained in detail in the sequence diagram in Fig.  \ref{fig5}.

In this scenario we used two drones (Drone 0 and Drone 1), both drones are displayed in a Gazebo simulator with similar physical characteristics, they perform a Diffie-Hellman key exchange and then exchange authenticated data using ROS. The process goes as follows:
\begin{enumerate}
\item Both drones generate Diffie-Hellman Parameters (the generator and the key size) using the cryptography library.
\item Both drones generate their private and public keys using the generated Diffie-Hellman parameters.
\item Drone 0 and Drone 1 initialize their ROS nodes which are called 'drone0\_authentication\_and\_send' \& 'drone1\_authentication\_and\_receive' respectively.
\item Drone 0 and Drone 1 log and publish their public keys to respectively the '/drone0/dh\_public\_key' \& the '/drone1/dh\_public\_key' ROS topics then wait for the other drone’s public key.
\item Drone 1 subscribes to the '/drone0/dh\_public\_key' ROS topic to receive the other drone's public key. It then performs a Diffie-Hellman key exchange and displays a message once the key exchange is complete.
\item Drone 0 subscribes to the '/drone1/dh\_public\_key' ROS topic to receive the other drone's public key. It then performs Diffie-Hellthe man key exchange and displays a message once the key exchange is complete.
\item Drone 1 subscribes to the '/drone1/authenticated\_data' ROS topic to receive authenticated data from Drone 0.
\item Drone 0 generates authenticated data and an HMAC using the shared key obtained from Diffie-Hellman key exchange.
\item Drone 0 publishes the authenticated data and HMAC to the '/drone1/authenticated\_data' ROS topic.
\item Drone 1 receives the authenticated data and verifies the HMAC.
\end{enumerate}
If drone 1 does not find a matching HMAC it displays an error and discards data. This communication flow ensures secure key exchange and data transfer between the two drones in a ROS environment. HMAC provides a mean for drones to authenticate each other by ensuring that the exchanged data has not been altered during the communication.

\section{Evaluation}
In this section, we identified the parameters  that could potentially influence our solution either positively or negatively.Subsequently, we evaluated these parameters to propose enhancements for future work. We used two different evaluation methods: the first based on the time taken for the process to finish, and the second involving a key tampering scenario to simulate man-in-the-middle attacks.

\subsection{Time-based Evaluation}
\begin{figure}[htbp]
\centerline{\includegraphics[width=\columnwidth]{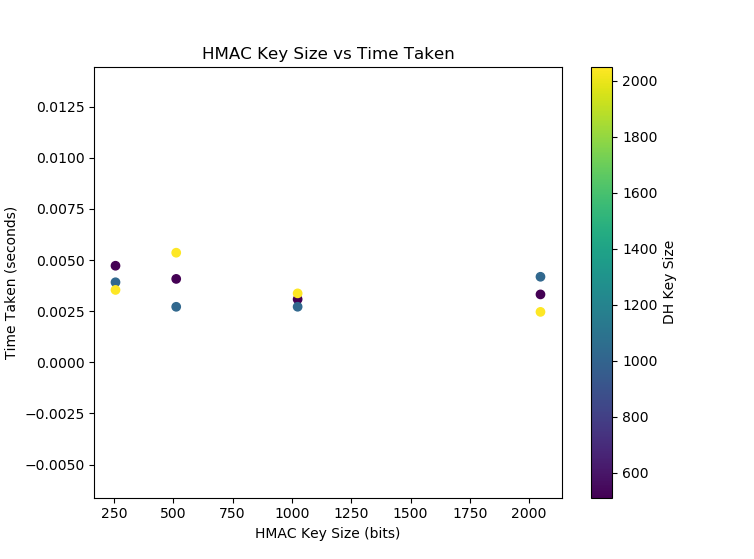}}
\caption{Scatter plot representing key variations against taken time.}
\label{fig6}
\end{figure}

\begin{figure}[htbp]
\centerline{\includegraphics [width=\columnwidth]{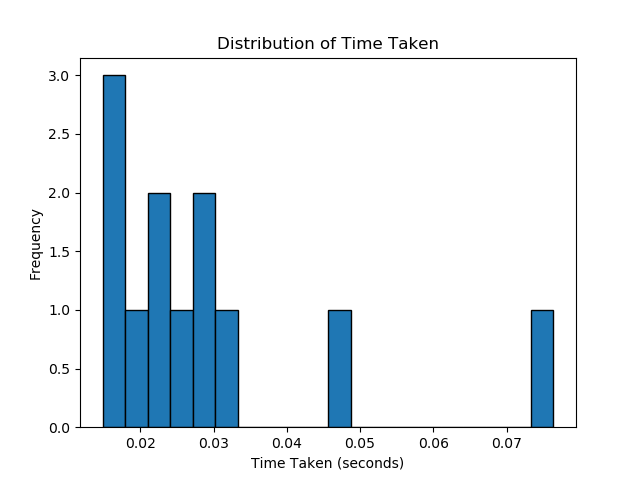}}
\caption{Histogram representing results of the time-based evaluation.}
\label{fig7}
\end{figure}

In this approach, we considered various values for both the Diffie-Hellman and HMAC keys. We conducted tests by varying the keys each time. The chosen values are the following:
For Diffie-Hellman: \[512, 256, 1024, 2028, 4096\]  \&  For HMAC: \[512, 1024, 2048\]
The test was conducted by repeating the key calculation, exchange, and authentication processes from start to finish using a different key each time. Each DH value is tested with all HMAC values until all DH values had been tested. Results were then illustrated in Figs. \ref{fig6} and \ref{fig7} respectively, representing a scatter plot and a histogram.

 From the representation, we observe that the Time Taken increases as the HMAC key size increases. However, it remains significantly below 1 or 2 seconds, which is a promising result considering the efficient resources utilized in the environment, contributing to the reduced time. During testing, when a larger DH key size, such as 4096, was employed, the program execution slowed considerably, taking over 10 minutes. Despite not obtaining a result with this key size, 2048 remains a highly suitable and secure choice, making it challenging for attackers to decrypt information once encrypted."

The scatter plot illustrates the relationship between HMAC Key Size, DH Key Size, and the Time Taken, with each point representing a combination of HMAC Key Size and DH Key Size. The color gradient indicates the DH Key Size, with lighter shades representing larger keys and darker shades indicating smaller key sizes. 

From the representation, we observe that the Time Taken increases as the HMAC key size increases. However, it remains significantly below 1 or 2 seconds, which is a promising result considering the efficient resources utilized in the environment. The latter, usually adds computing complexity thus the Time taken to finish the process which did not happen in our scenario. However, when we reached higher values, such as DH key sizes equal to or higher than 4096, the program execution slowed significantly, taking over 10 minutes. Despite not obtaining results with these higher values, 2048 remains a highly suitable and secure key size. It effectively complicates attackers' attempts to decrypt encrypted information.

The second illustration presents the results of another series of tests conducted with the same parameters. We reran the program to ensure that results are consistent and have not been altered. The distribution of Time Taken across all test cases reveals that the majority fall within a specific range of time while cases with exceptionally high or low Time Taken are identified as outliners. This suggest that the overall results positively impact performance, as the Time Taken remains within the range previously depicted in the scatter plot.

\subsection{Tampering Evaluation}
It is well known that Man in the Middle attacks significantly compromise data exchange security.  despite being a secure algorithm, the Diffie-Hellman key exchange lacks the ability to verify the integrity of exchanged data, making it vulnerable to such attacks. Incorporating the HMAC algorithm has bolstered authentication and added data integrity to our system. However, during the public key exchange phase, attackers can still intercept and manipulate the keys before injecting them back into the communication channel. Thus, UAVs must possess the capability to detect tampering to ensure data authenticity.

\begin{figure}[htbp]
\centerline{\includegraphics[width=\columnwidth]{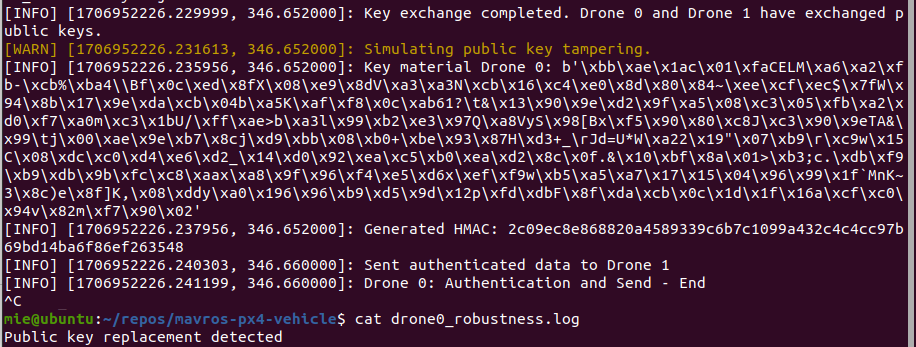}}
\caption{Key Tampering \& key replacement for Drone 0.}
\label{fig8}
\end{figure}

\begin{figure}[htbp]
\centerline{\includegraphics[width=\columnwidth]{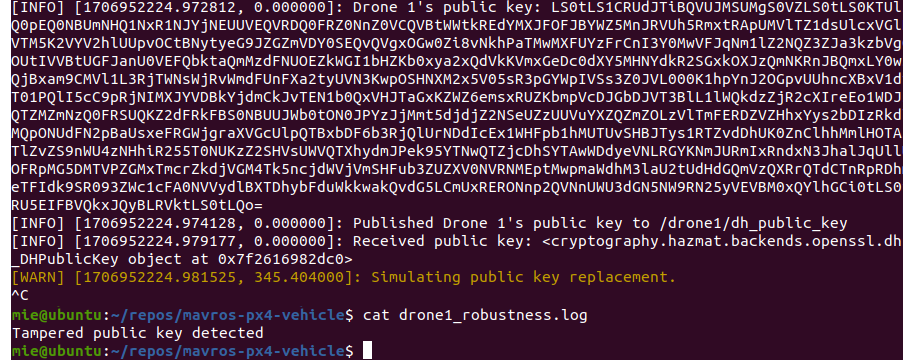}}
\caption{Key Tampering \& Key Replacement for Drone 1.}
\label{fig9}
\end{figure}

For testing and evaluation, we simulated a Man in the Middle attack by targeting the exchanged public keys. Our approach was simple and straightforward: we implemented two functions for each Drone0 and Drone1. The first function simulates tampering by altering a portion of the key, while the second function simulates key replacement.To determine the detection process, we incorporated a function that randomly decides whether to simulate tampering or key replacement.Both drones have this function, and each independently makes its own random decision. Consequently, the outcome could vary: both drones may detect tampering, both may detect key replacement, or one may detect tampering while the other detects key replacement.
The results were recorded in a log file for each drone, as depicted in Figs. \ref{fig8} and \ref{fig9} for Drone 0 and Drone 1, respectively.

As demonstrated above, the tests yielded successful results; both drones detected alterations and manipulations, showcasing the algorithm's effectiveness in guarding against Man-in-the-Middle attacks.

\section{Conclusion}
The application of security and reliability in UAV networks presents a huge challenge for researchers, especially as when these integrate with futuristic 6G communication networks operating in complex environments. Thus, novel data exchange mechanisms and protocols are required to ensure efficiency, reliability and security. In this paper, we proposed a secure mechanism for inter-UAV communication to prevent attacks while maintaining resilience in untrusted environments and being flexible. We proposed the combination of the Diffie-Hellman key exchange algorithm and HMAC for UAV authentication and ensuring data integrity. We conducted two testing scenarios to asses the algorithm's performance and its limits.The first test focused on evaluating the processing time by varying key sizes, with results showing that processing time remained below one second and exceeded 10 minutes for for sizes above 4096. In the second test, we simulated a Man-in-the-Middle attack by incorporating key replacement, tampering, and detection functions. Results demonstrated the system's scalability and resilience, indicating potential for additional security implementations to further enhance its robustness.
\section{Perspectives \& Future Works}
Our work aims  to achieve optimal reliability for UAV communications, particularly in the context of futuristic 6G networks that handle vast and complex amounts of data. To maximize and facilitate UAV integration, we are exploring novel paradigms and innovative approaches to leverage existing and emerging technologies.
Future works will focus on:
\begin{itemize}
\item Drone energy consumption. 
\item Latency in drone communication. 
\item Implementing reliable communication approaches for a swarm of drones.
\item Using cloud computing to reduce drone overhead and resource consumption.
\end{itemize}

\end{document}